\documentclass[twocolumn,10pt]{asme2ej}
\usepackage{amsmath,amssymb,amsfonts,amsthm}
\usepackage[linesnumbered,ruled,vlined]{algorithm2e}
\usepackage{graphicx}
\usepackage{multirow}
\usepackage[flushleft]{threeparttable}
\usepackage[dvipsnames,usenames]{color}
\usepackage{url}
\usepackage{diagbox}
\usepackage{gensymb}
\graphicspath{{figures/}}

\usepackage{fancyhdr}
\newtheorem{proposition}{Proposition}
\newtheorem{remark}{Remark}
\def\cotwo{CO$_2$}

\def\dist{\ensuremath{q^\mathrm{int}}}
\def\qhvac{\ensuremath{q^\mathrm{hvac}}}
\def\prh{\ensuremath{P^\mathrm{rh}}}
\def\pcc{\ensuremath{P^\mathrm{cc}}}

\def\qfan{\ensuremath{P^\mathrm{fan}}}
\def\cop{\ensuremath{COP}}
\def\solirr{\ensuremath{\eta^\mathrm{sol}}}
\def\Ae{\ensuremath{A_\mathrm{e}}}
\def\mdot{\ensuremath{\dot{m}}}
\def\cpa{\ensuremath{C_\mathrm{pa}}}

\def\tz{\ensuremath{T^\mathrm{z}}}
\def\tw{\ensuremath{T^\mathrm{w}}}
\def\tca{\ensuremath{T^\mathrm{ca}}}
\def\tsa{\ensuremath{T^\mathrm{sa}}}
\def\tma{\ensuremath{T^\mathrm{ma}}}
\def\toa{\ensuremath{T^\mathrm{oa}}}
\def\tsamin{\ensuremath{T^\mathrm{sa,min}}}
\def\tsamax{\ensuremath{T^\mathrm{sa,max}}}
\def\tsarate{\ensuremath{T^\mathrm{sa,rate}}}
\def\tzmin{\ensuremath{T^\mathrm{z,min}}}
\def\tzmax{\ensuremath{T^\mathrm{z,max}}}

\def\tp{\ensuremath{N}}
\def\tc{\ensuremath{N_\mathrm{c}}}

\def\MATLAB{\text{MATLAB}\ensuremath{^\copyright}}
\def\simulink{\text{SIMULINK}\ensuremath{^\copyright}}
\def\cvx{\text{CVX}\ensuremath{^\copyright}}
\def\ipout{\text{Ipopt}\ensuremath{^\copyright}}
\def\eqdef{\ensuremath{:=}}
\def\R{\mathbb{R}}

\def\S{\mathbb{S}}

\newcommand{\Set}{\mathbf{S}}

\def\eqdef{\ensuremath{:=}}

\def\pb#1{\footnote{{\color{blue}{PB}: } #1}}

\newcommand{\Lagr}{\mathcal{L}}

\newcommand{\Z}{\mathbb{Z}}

\usepackage{xparse}

\ExplSyntaxOn
\cs_new:Nn \pb_prop_gset_bykeys:Nn
{
	\prop_clear:N \l__pb_temp_prop
	\keys_set:nn { pb/propbykey } { #2 }
	\prop_gset_eq:NN #1 \l__pb_temp_prop
}
\cs_generate_variant:Nn \pb_prop_gset_bykeys:Nn { c }

\keys_define:nn { pb/propbykey }
{
	unknown .code:n = \prop_put:Nxn \l__pb_temp_prop { \l_keys_key_tl } { #1 }
}

\cs_generate_variant:Nn \prop_put:Nnn { Nx }

\NewDocumentCommand{\setupcollaborator}{mm}
{
	\prop_new:c { g_collaborator_#1_prop }
	\pb_prop_gset_bykeys:cn { g_collaborator_#1_prop } { #2 }
}

\NewDocumentCommand{\selectcollaborator}{m}
{
	\prop_map_inline:cn { g_collaborator_#1_prop }
	{
		\tl_set:cn { ##1 } { ##2 }
	}
}
\ExplSyntaxOff

\setupcollaborator{PBmac}
{
	NAME=Prabir Mac , laptop or mac mini,
	DiCEbibPATH=/Users/pbarooah/Dropbox/Dropbox/DiCElab_bib,
	ACbibPATH=/Users/pbarooah/Dropbox/Dropbox/austin_bib,
	JBbibPATH=/Users/pbarooah/Dropbox/Dropbox/bibs,
	TZbibPATH = /Users/pbarooah/Dropbox/Dropbox/TTbib,
	figPATH= ./figures,
}
\setupcollaborator{TZlinux}
{
	NAME=TZ at lab,
	DiCEbibPATH=../../bib/DiCElab_bib,
	ACbibPATH=../../bib/austin_bib,
	JBbibPATH=../../bib/bibs,
	TZbibPATH = ../../bib/TTbib,
	figPATH= ./,
}
\setupcollaborator{TZmac}
{
	NAME=TZ at her macbook,
	DiCEbibPATH=../../bib/DiCElab_bib,
	ACbibPATH=../../bib/austin_bib,
	JBbibPATH=../../bib/bibs,
	TZbibPATH = ../../bib/TTbib,
	figPATH=/Users/ZTT/Desktop/Dropbox (UFL)/Papers/19_Zeng_AutonomousMPC_journal/figures, 
}

\selectcollaborator{TZlinux} 

\title{An adaptive MPC scheme for energy-efficient control of building HVAC systems}

\author{Tingting Zeng and Prabir Barooah\thanks{This research is partially supported by NSF through grant 1463316 and grant 1934322. Corresponding author: Tingting Zeng. Email: tingtingzeng@ufl.edu.}
    \affiliation{
	University of Florida\\
	Gainesville, Florida USA\\
    }	
}

\begin{document}

\maketitle    

\begin{abstract}
{\it 
\textbf{Abstract:} An autonomous adaptive MPC architecture is presented for control of heating, ventilation and air condition (HVAC) systems to maintain indoor temperature while reducing energy use. Although equipment use and occupant changes with time, existing MPC methods are not capable of automatically relearning models and computing control decisions reliably for extended periods without intervention from a human expert. We seek to address this weakness. Two major features are embedded in the proposed architecture to enable autonomy: (i) a system identification algorithm from our prior work that periodically re-learns building dynamics and unmeasured internal heat loads from data without requiring re-tuning by experts. The estimated model is guaranteed to be stable and has desirable physical properties irrespective of the data; (ii) an MPC planner with a convex approximation of the original nonconvex problem. The planner uses a descent and convergent method, with the underlying optimization problem being feasible and convex. 
A year long simulation with a realistic plant shows that both of the features of the proposed architecture - periodic model and disturbance update and convexification of the planning problem - are essential to get the performance improvement over a commonly used baseline controller. Without these features, though MPC can outperform the baseline controller in certain situations, the benefits may not be substantial enough to warrant the investment in MPC.}
\end{abstract}

\section{Introduction}
Heating, ventilation, and air conditioning (HVAC) systems are responsible for approximately 40\% of the total energy consumption of buildings in the USA~\cite{energyoutlook11_eia:18}. It has been recognized by many researchers that instead of the traditional rule-based control systems, an optimization based controller - especially Model Predictive Control (MPC) - is a highly promising approach to reduce energy use; see the review papers~\cite{afram2014theory,serale2018model}.

In spite of extensive studies and even successful demonstration projects~\cite{castilla2014thermal,vsiroky2011experimental}, MPC has not been widely adopted in practice. The bottlenecks - which have been discussed extensively as well - can be summarized into \emph{lack of autonomy} of existing control architectures that use MPC. By \emph{autonomous MPC} we mean an MPC scheme capable of reliably computing high-quality control decisions at all times without the need for human intervention. A building's and its equipment's behavior are quite complex and uncertain, so the models needed by MPC need to be learned from data~\cite{smarra2018data,Foucquier2013state}. Since the building's behavior also changes with time - albeit slowly - the models need to be updated over time~\cite{smarra2018data}. The overall architecture thus needs to be adaptive.

Although there is an extensive literature on identification of HVAC system models from data, the vast majority of the existing methods cannot be used for autonomous adaptation. These algorithms involve solving a non-convex optimization problem with few guarantees on the quality of the model fit~\cite{kelman2011bilinear,razmara2015optimal}. Depending on the type and quality of data used, they require re-tuning of hyper-parameters by a human expert. Clearly such an approach cannot lead to an autonomous control system. A related issue that the unmeasurable internal heat gains from occupants is substantial, while most identification methods ignore them~\cite{Oldewurtel2012use,bualan2011parameter,privara2011model}, which can lead to poor quality models. 


The planning problem that MPC solves at every decision instant to compute control commands should be feasible and convex. With a nonconvex problem the planner can fail to converge to a local minimum within the allowed computation time. Infeasibility has the same effect. In either case, a rule-based controller must be used as back up when the non-convex planner cannot provide a control command. Switching between controllers can cause poor performance. The MPC planning problem is usually non-convex due to bilinearities in models and cost functions~\cite{smarra2018data,kelman2011bilinear,ma2015stochastic,parisio2014control,atam2015convex}. Most works on HVAC MPC ignore the issue of reliability of the decisions computed by a non-convex planner, especially over long periods of operation.  

\begin{figure}[htb]
	\centering	\includegraphics[width=0.8\linewidth]{\figPATH/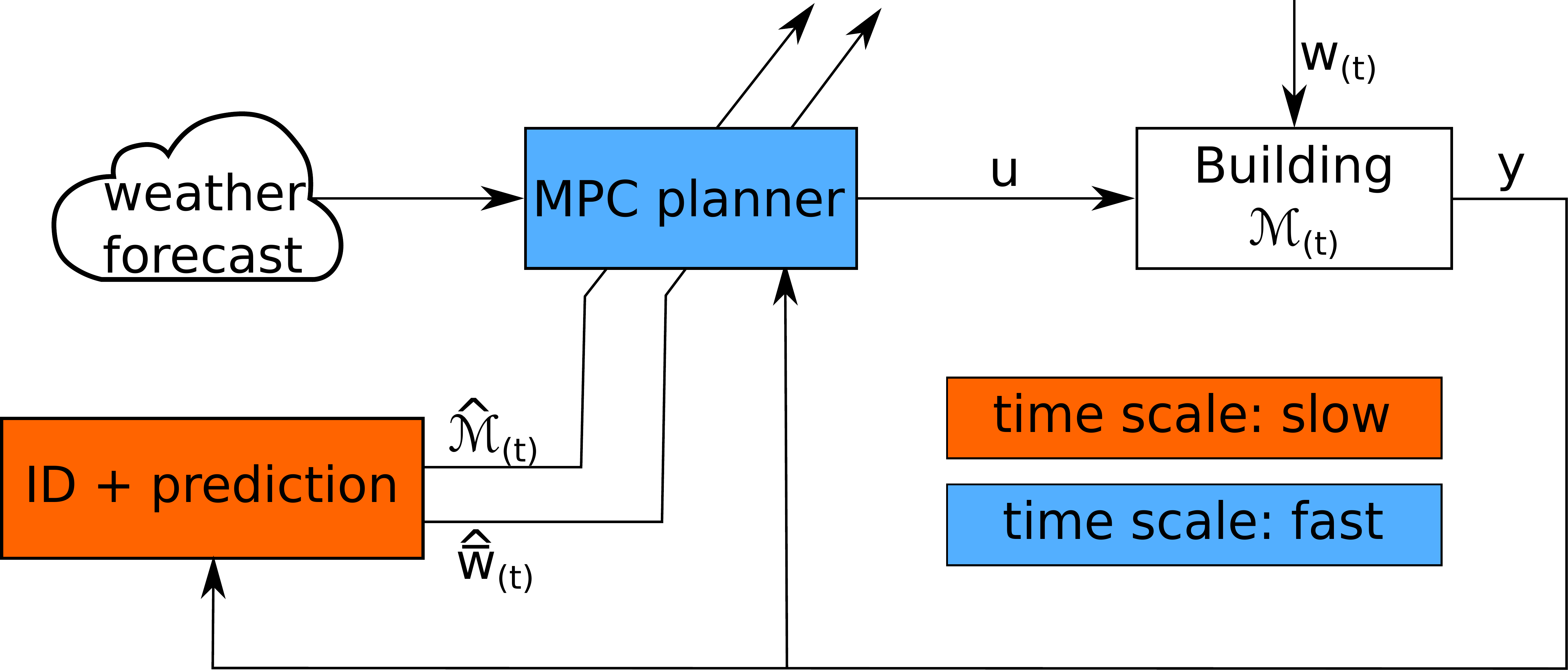}
	\caption{Structure of the desired building adaptive MPC architecture.}
	\label{fig:MPCflowBrief}
\end{figure}

In this paper we propose an adaptive MPC architecture for HVAC systems, shown in Fig.~\ref{fig:MPCflowBrief}, that is capable of operating autonomously for long periods of time without intervention of a human expert.
The ``ID + prediction'' block uses an algorithm proposed in our prior work~\cite{ZengSimultaneousAutomatica:2018} to identify the plant model  and the unmeasured internal disturbance from easily measured input-output data. This algorithm involves solving an optimization problem that is always feasible and convex, and the model it identifies ($\hat{\mathcal{M}}$) is guaranteed to be stable and possess properties that are consistent with properties of a building HVAC system. The algorithm has one hyper parameter that needs to be tuned only once. In short, the identification algorithm does not need any human intervention when new data is fed into it periodically, say, every week, to update the model. The past disturbance identified by the algorithm is used to forecast the future disturbance ($\hat{\bar{w}}$) that is in turn used by the MPC planner. 

The ``MPC planner'' block of the system uses the model and disturbance forecasts to decide control commands so as to maintain indoor climate while reducing energy use. We provide a convex approximation algorithm to approximately solve the nominal non-convex planning problem. We show that the algorithm is a feasible, descent, and convergent algorithm. Thus the MPC planner block can compute decisions autonomously without human expert. We also show that among the many ways of convexifying these types of non-convex problems, the proposed approach is the only one applicable to our specific problem structure.

The proposed convex planner and the associated analysis is the first novel contribution of the paper. The second contribution is the performance assessment of the closed-loop system for a year-long period. Numerical results show that the proposed MPC scheme is not only more energy efficient and better at indoor climate control than a conventional baseline controller. More importantly, these simulations show that both features of the proposed design - periodic update of the model and disturbance and convexified MPC planner - are necessary to get the performance improvement over the baseline controller. These discoveries are made possible only due to the long duration for which simulations are conducted. While that is perhaps not surprising for the role of periodic model update, the discovery on the role of convexification also required the year-long simulation. In particular, the non-convex controller was seen to perform as well as the convex one in all but a few rare instances. In those few rare cases, however, the performance of the non-convex planner was catastrophic.

The article makes three additional contributions over the preliminary version~\cite{zeng2020autonomous}. (i) We provide new analysis regarding the appropriate convexification method for the MPC planner, while~\cite{zeng2020autonomous} did not have any such results. (ii) This article includes closed loop simulations for a year long period while the preliminary version only included three weeks of simulation. (iii) Comparison with three additional architectures, each obtained by removing model update or convexification or both, are provided. These comparisons reveal which of these features of the proposal are useful or necessary (or not).

The rest of this paper is organized as follows. Section~\ref{sec:lit} provides a review of relevant work on HVAC MPC. Section~\ref{sec:arch} describes the HVAC control problem. Sections~\ref{sec:MPCplanner} and~\ref{sec:ID+prediction} present the components of ``MPC planner" and ``ID + prediction" respectively. The simulation is set up in Section~\ref{sec:setup}, and the results are presented in Section~\ref{sec:results}. Finally Section~\ref{sec:conclusion} concludes this work. 

\subsection{Literature review}\label{sec:lit}
Dynamic models of building HVAC systems are typically nonlinear, which makes the planning problem in MPC a non-convex optimization problem. The nonlinearity comes from the existence of the bilinear terms - products between a state (temperature) and a control command (airflow rate). Sometimes the dynamic model is linearized to obtain a convex problem. Among works adopting this approach, some assume the value for a certain control command is known so that the product in which that command appears becomes linear in the remaining decision variables~\cite{aswani2011reducing,parisio2014control,razmara2015optimal}. This reduces a degree of freedom that MPC can use. 
Others linearize around a trajectory, which requires an optimal (or at least a near optimal) trajectory first~\cite{maasoumy2014handling,ma2015stochastic,koehler2013building}. The quality of a linearized model is sensitive to the choice of the trajectory, and determining such a trajectory is challenging. After all, if it were easy there would be little need for MPC. Identifying a linear black box model directly from data is also not straightforward (we discuss this later in Section~\ref{sec:MPCplanner}). Recent progress in this direction is made in~\cite{ZengSimultaneousHPB:2018,ZengSimultaneousAutomatica:2018}, which identifies a linear model in which the input is the heat gain due to the HVAC system. However, the model is still not linear with respect to the control commands such as air flow rate. Convex relaxation of the MPC planning problem is thus far from trivial.


There has been recent attempts at convexification of the non-convex planning problems encountered in HVAC MPC~\cite{Oldewurtel2012use,kelman2011bilinear,atam2015convex}. Ref~\cite{Oldewurtel2012use} does not require constraints to be satisfied at all time, but only with a pre-defined probability. Therefore the resulting solution may not satisfy actuator constraints. In~\cite{kelman2011bilinear}, values of the Lagrange multipliers are required for reformulation of the problem. The convexification approach using a McCormick envelope considered in~\cite{atam2015convex} requires feasibility of the original problem (without slack) for all time. The original problem is likely to be infeasible when disturbance is large, since the actuator limits will prevent them from maintaining state constraints.

Another particular challenge is that the disturbance is also a large part of the heat load and hence a large part of the energy consumption in buildings. It is unlikely to be negligible, especially for large buildings such as the auditorium considered as the testbed in our study. Therefore disturbance prediction is needed to achieve the promised performance of MPC. However most of the prior works ignore the effects of the disturbance~\cite{Oldewurtel2012use,bualan2011parameter,privara2011model}, which may lead to erroneous decision. Some estimate the disturbance based on real-time occupancy recognition or equipment use measurements~\cite{Aftab2017automatic,chang2013statistical}, which requires additional sensing. Some assume perfect knowledge of future disturbance~\cite{RamanMPC_AE:2020,Hou2017distributed,kelman2011bilinear,atam2015convex}, but such knowledge is not available during implementation. 

\section{Architecture}\label{sec:arch}
\subsection{Problem description}
\begin{figure}[htb]
	\centering
	\includegraphics[width=1\linewidth]{\figPATH/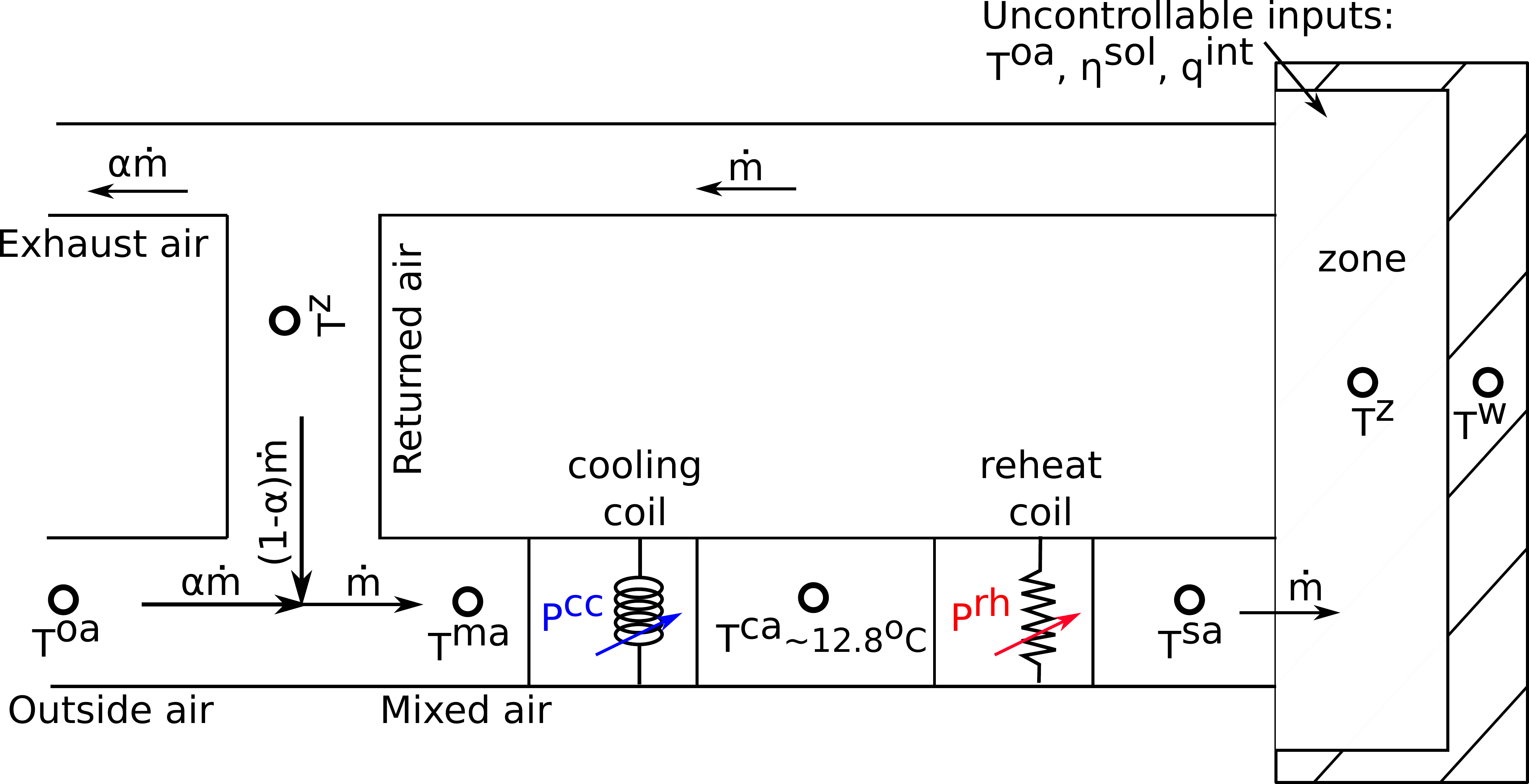}
	\caption{A schematic of a typical single-zone variable air volume HVAC system used in a commercial building.}
	\label{fig:HVAC}
\end{figure}

The focus of this paper is the indoor climate control of a single-zone HVAC system shown in Figure~\ref{fig:HVAC}.
In such a system, part of the air exhausted from the zone is recirculated and then mixed with outdoor air (OA) at a specified ratio. This mixed air (MA) is usually warm and humid, especially for hot-humid climates, and is therefore cooled and dehumidified by passing through a cooling coil. Dehumidification requires that the air is cooled enough for the water vapor to condense out of the air stream, so the conditioned air (CA) temperature (after the cooling coil), is usually too cold for a comfortable indoor climate. It is reheated by the reheat coil up to supply air (SA) temperature, and then delivered into the zone. 

The goal of the control system designed in this study is to decide the control commands to maintain the zone temperature (\tz) within time-varying pre-determined bounds, while keeping the energy use as small as possible. The control commands are the setpoints for total airflow rate (\mdot) and  supply air temperature (\tsa). Lower level PI controllers will maintain these setpoints by varying fan speed and reheat valve position. 

Although conditioned air temperature $\tca$ and the outdoor air ratio $\alpha$ (ratio of outdoor air flow rate to supply air flow rate) can also be varied, in this paper we assume they are fixed. The conditioned air temperature is typically set to $12.8$\degree C in order to maintain zone humidity, which is an important aspect of thermal comfort~\cite{ashrae55:2017}. Similarly, the outdoor air ratio $\alpha$ is pre-specified at a constant value, and the minimum allowed value for the supply air flow rate \mdot\ is computed so that the OA flow rate $\alpha \mdot$ meets ventilation requirements~\cite{ASHRAE62-2016}.

\subsection{Control system architecture}
The control architecture proposed in this paper is shown in Figure~\ref{fig:MPCflowBrief}. It involves two main components: (i) the \emph{ID and prediction} block, and (ii) \emph{MPC planner} block that uses the models and forecasts to compute control commands.

Model predictive control of a system $x_{k+1}=f_k(x_k,u_k,v_k)$, $y _k = h_k(x_k,u_k,v_k)$, with $x$ being the state, $u$ being the control command and $v$ being the uncontrollable inputs, involves minimization of a cost function $J_i = \sum_{k=i}^{i+N-1}c_k(\hat{x}_{k+1},u_k,\hat{v}_k)$ over the planning horizon $N$ with $c_k(\cdot)$ being the energy used during the interval between $k$ and $k+1$. At time index $i$, an optimization problem of minimizing $J_i$ subject to the system model and other constraints is posed based on the current estimate $\hat{x}_i$ of the state $x_i$ and forecasts $\hat{v}$ of uncontrollable inputs $v$. The solution to this problem yields optimal commands $u_i,u_{i+1},\dots,u_{i+N-1}$. The first entry, $u_i$, is implemented. At the next time index $i+1$, the procedure is repeated.

In the adaptive architecture proposed here, the model is updated periodically by the system identifier, though at a much slower time scale than that of the control command update. In the numerical studies later reported, the model is updated every week while control commands are updated every fifteen minutes.

\section{(Block II) MPC planner}\label{sec:MPCplanner}
We start with the second block (MPC planner) of Fig.~\ref{fig:MPCflowBrief}, since it is in charge of computing control commands, and the other blocks are there merely to support it. The MPC planner needs models to describe the energy consumption and the temperature dynamics, which appear as part of the cost and constraints in the planning problem. Energy is the integral of power, and the power consumption of the HVAC system consists of fan power, cooling power and reheat power. The fan power is modeled as~\cite{RamanAnalysisHPB:2018}:
\begin{align}\label{eq:qfan_def}
\qfan_k = a_f\mdot_k^2,
\end{align}
with \mdot\ being the total airflow rate (kg/s). The cooling power $\pcc_k$ is the electrical power consumed by the chiller to cool down the warm mixed air as it passes through the cooling coil:
\begin{align}\label{eq:qcc_def}
\pcc_k = \begin{cases}
\frac{\cpa\mdot_k(\tma_k-\tca)}{\cop} & \tma_k > \tca\\
0 & \text{otherwise}
\end{cases},
\end{align}
where \cpa\ is the specific heat of air at constant pressure, \tca\ (\degree C) is the conditioned air temperature, \cop\ is the chiller performance coefficient, and the mixed air temperature \tma\ (\degree C) is given by:
\begin{align}\label{eq:tma_def}
\tma_k = \alpha\toa_k+(1-\alpha)\tz_{k},
\end{align}
where $\alpha$ ($=\frac{\mdot^{oa}}{\mdot}$) is the outdoor air ratio, \toa\ is the outdoor air temperature (\degree C), $\mdot^{oa}$ is the outdoor airflow rate (kg/s), and \tz\ (\degree C) is the indoor zone temperature.  The power consumed by reheat coil is modeled as the heat it adds to the conditioned air:
\begin{align}\label{eq:qrh_def}
\prh_k & = \cpa\mdot_k(\tsa_k-\tca),
\end{align}
where \tsa\ is the supply air temperature (\degree C).

Dynamics of zone temperature is modeled by the following second order linear system with one output (indoor zone temperature $\tz$), and four inputs (heat injected by the HVAC system $\qhvac$, outdoor temperature \toa ($\degree$C), solar irradiance \solirr (kW/m$^2$), transformed disturbance $\bar{w}$): 
\begin{equation}\label{eq:thermal_mdl}
\begin{split}
x_{k+1} &= A x_k + B\qhvac + F v_k\\
\tz_k   &= C x_k + D\qhvac + G v_k    
\end{split}
\end{equation}
where $x\in \R^2$ is the state and $A\in \R^{2\times 2}$, $B\in \R^{2\times 1}$, $C\in \R^{1\times 2}$, $D\in \R$, $F\in \R^{2\times 3}$, $G\in \R^{1 \times 3}$ are appropriate system matrices. The four inputs are separated into the single controllable input $\qhvac$ and three uncontrollable inputs $v\eqdef [\toa,\solirr,\bar{w}]^T\in \R^3$, where $\bar{w}$ is the transformed version of the internal heat load \dist\ (kW); see~\cite{ZengSimultaneousAutomatica:2018} for details. It captures the effect of \dist\  on the zone temperature $\tz$. The quantity \qhvac\ is related to the two actuation commands (supply airflow rate \mdot\ and the deviation of supply air temperature \tsa) via the bilinear relation:
\begin{align}\label{eq:qhvac_def}
\qhvac_k=\cpa\mdot_k(\tsa_k-\tz_k).
\end{align}

\begin{remark}
	Although \qhvac\ is considered the controllable input in~\eqref{eq:thermal_mdl}, it cannot be commanded directly. Only \mdot\ and \tsa\ can be commanded. Treating \qhvac\ as the controllable input helps in two ways. Firstly, it makes the model~\eqref{eq:thermal_mdl} linear, which aids model identification (discussed in Section~\ref{sec:ID+prediction}). Secondly, the linear model is a convex constraint in the optimization problem the planner has to solve. We emphasize that a linear model structure with \mdot\ and \tsa\ as inputs, even though conceptually possible, is not useful for eventual use in MPC. The reason is that the sign of the DC gain (from \mdot\ to \tz) depends on whether the control commands are having a cooling or heating effect on the zone. If the supply air temperature \tsa\ is higher than the zone temperature \tz, increasing \mdot\ will increase the zone temperature. So the DC gain is positive in such a scenario. The opposite happens when \tsa\ is lower than \tz. Now the DC gain has to be negative. However a-priori knowledge of whether the control inputs will lead to heating or cooling is not available since that depends on both the state an control command. 
\end{remark}

\subsection{Nominal non-convex planner}\label{subsec:noncvx_ctrl}
The goal of the MPC planner is to compute the control commands over the planning horizon, supply airflow rate \mdot\ and supply air temperature \tsa, to maintain thermal comfort while reducing energy use over that horizon. A direct translation of this goal into an optimization problem will be a non-convex problem, partly due to the bilinearity in~\eqref{eq:qhvac_def}. We first present this problem below, and then use it as a stepping stone to formulate a convex approximation that is actually used in the proposed MPC planner.

For notational simplicity, the current time index $i$ is assumed to be 0 in this section. Define the decision variables as $z_k\eqdef [\mdot_k,\tsa_k,\tma_k,\tz_k,\qhvac_k,x^T_{k+1},\epsilon^{min}_k,\epsilon^{max}_k]^T \in \R^{9}$, in which $x \in \R^2$ is the state of the thermal model~\eqref{eq:thermal_mdl}, and \mdot\ and \tsa\ are the control commands, and \tp\ is the planning horizon. Let $\hat{x}_0$ be the estimate of the current state obtained from a state estimator, and let $\hat{v}_k$ ($\eqdef [\hat{T}^{oa}_k,\hat{\eta}^{sol}_k,\hat{\bar{w}}_k]^T$) be the prediction of the uncontrollable inputs, for $k=0,\dots,\tp-1$. Specifically, $\hat{T}^{oa}$ and $\hat{\eta}^{sol}$ are from weather forecast, and $\hat{\bar{w}}$ is provided by a disturbance predictor which will be discuss later in Sec.~\ref{subsec:ctrl_arch}.

The \textbf{nominal non-convex planning problem} is:
\begin{subequations}\label{eq:opt_org}
	\begin{align}
	\min_{z_k|_{k=0}^{\tp-1}} J, \quad & J \eqdef \sum_{k=0}^{\tp-1}\big(\Delta t(\prh_k+ \pcc_k + \qfan_k) +\rho(\epsilon^{min}_k+\epsilon^{max}_k)\big) \tag{\ref{eq:opt_org}}\\
	\text{ s. t. }
	-&\qhvac_k+\frac{1}{2}z_k^TP_cz_k=0 \label{eq:opt_qhvac} \\ 
	-&x_{k+1} + A x_k + B\qhvac_k + F \hat{v}_k=0_{2\times 1} \label{eq:opt_ss1},\quad x_0 = \hat{x}_0 \\ 
	-&\tz_k + C x_k + D\qhvac_k + G \hat{v}_k=0 \label{eq:opt_ss2} \\ 
	-&\tma_k +(1-\alpha)\tz_{k}+ \alpha\toa_k = 0 \label{eq:opt_tma} \\ 
	-&\mdot_k\leq -\mdot^{min}, \quad \mdot_k\leq  \mdot^{max} \label{eq:opt_mdot} \\ 
	-&\mdot_{k+1}+\mdot_{k}\leq \mdot^{rate}\Delta t \label{eq:mdot_rate1}\\ 
	&\mdot_{k+1}-\mdot_{k}\leq \mdot^{rate}\Delta t \label{eq:mdot_rate2}\\ 
	-&\tsa_k\leq -{\tsamin}, \quad \tsa_k\leq \tsamax \label{eq:opt_tsa} \\ 
	-&\tsa_{k+1}+\tsa_{k}\leq \tsarate \Delta t \label{eq:tsa_rate1} \\ 
	&\tsa_{k+1}-\tsa_{k}\leq \tsarate \Delta t \label{eq:tsa_rate2} \\ 
	-&\tz_k-\epsilon^{min}_k\leq -\tzmin \label{eq:opt_tz1} \\ 
	&\tz_k-\epsilon^{max}_k\leq \tzmax \label{eq:opt_tz2} \\ 
	-&\epsilon^{min}_k\leq 0, \quad -\epsilon^{max}_k\leq 0 \label{eq:opt_tzslack} \\ 
	& k=0,\dots,\tp-1, \nonumber 
	\end{align}
\end{subequations}
where
\begin{align}
\begin{aligned}\label{eq:qhvac_quad}
P_c & =
\left[
\begin{array}{c|cccc}
0 & \cpa & 0 & -\cpa & 0_{1\times 5} \\ \hline 
\cpa & & & & \\ 
0 & & \raisebox{0pt}{$0_{8 \times 8}$} & &\\
-\cpa & & & &\\
0_{5\times 1} & & & &
\end{array}
\right],
\end{aligned}
\end{align}
and is obtained by rewriting~\eqref{eq:qhvac_def}.

Actuator constraints $[\mdot^{min},\mdot^{max}]$ and $[\tsamin,\tsamax]$, represent the lower and upper bounds of the airflow rate and the supply air temperature, respectively. The minimum supply airflow rate, $\mdot^{min}$, is computed based on ventilation requirements~\cite{ASHRAE62-2016}. To ensure reheat coil can only add heat, we require $\tsamin=\tca$. Thermal comfort bounds are $[\tzmin, \tzmax]$. Slack variables $\epsilon^{min},\epsilon^{max}$ are used to relax the thermal the thermal comfort bounds from a fixed range $[\tzmin, \tzmax]$ to a variable range   $[\tzmin-\epsilon^{min}, \tzmax+\epsilon^{max}]$. These slack variables help ensure that the problem is feasible. A high penalty parameter $\rho$ encourages the slacks variables to be small so that temperature violation - when it occurs - is small.

For later convenience, we note that $J$ can be compactly expressed as
\begin{align}
J	& =\sum_{k=0}^{\tp-1} \big(\frac{1}{2}z_k^TPz_k+ q^Tz_k \big), \label{eq:cost_quad}
\end{align}
where
\begin{align}
P& =  \left[
\begin{array}{c|ccc}
2\alpha_f & \cpa &\frac{\cpa}{\cop} & 0_{1\times 6} \\ \hline 
\cpa  & & & \\ 
\frac{\cpa}{\cop} & &\raisebox{0pt}{$0_{8 \times 8}$}& \\
0_{6\times 1} & & &
\end{array}
\right]\Delta t, \label{eq:opt_Porg} \\
q& = 
\begin{bmatrix}
-\cpa\tca\frac{1+\cop}{\cop}\Delta t & 0_{6\times 1} & \rho & \rho
\end{bmatrix},  \label{eq:opt_qorg}
\end{align}
by rewriting~\eqref{eq:qfan_def}-\eqref{eq:qcc_def} and~\eqref{eq:qrh_def}.

\begin{proposition}\label{prop:opt_org_feasible}
	Problem~\eqref{eq:opt_org} is feasible.
\end{proposition}
The proof of Proposition~\ref{prop:opt_org_feasible} is provided in the Appendix.

\subsection{Proposed convex planner}\label{subsec:cvx_ctrl}
The optimization problem~\eqref{eq:opt_org} is non-convex since the equality constraint~\eqref{eq:opt_qhvac} is bilinear, and the quadratic term in the cost~\eqref{eq:cost_quad} involves the indefinite matrix $P$. \textit{The goal now is to approximate the problem~\eqref{eq:opt_org} with a  convex problem, so that the approximation is easy  to solve and the obtained solution provides good approximation to that of problem~\eqref{eq:opt_org}.} 

The algorithm we propose to this end is described in Algorithm~\ref{algo:convexCtrl}. It uses the Convex-Concave Procedure (CCP)~\cite{lipp2016variations}. In Algo.~\ref{algo:convexCtrl}, the following terminology is used. Let $P=Q(\Lambda^+ +\Lambda^-)Q^T$ be the eigen-decomposition of the real symmetric matrix $P$ from Eq.~\eqref{eq:opt_Porg}, where $\Lambda^+\succcurlyeq 0$ is the positive semi definite part and $\Lambda^-\prec 0$ is the negative definite part. Define $P^+\eqdef Q\Lambda^+Q^T$ and $P^-\eqdef Q\Lambda^-Q^T$.
\begin{algorithm}[h]\label{algo:convexCtrl}
	\LinesNotNumbered
	\caption{Convex planner}
	\SetAlgoLined
	\KwIn{Initial guess $\zeta(0)$.} 
	$n\leftarrow 0$. \\
	\Repeat{$\|\zeta(n)-\zeta(n-1)\|\leq \delta $}{
		\textbf{Convexify:}	Form:
		\begin{align}\label{eq:qhvac_hat}
		\begin{aligned}
		&\hat{J} =  \sum_{k=0}^{\tp-1}\big(\frac{1}{2}z_k^TP^+z_k 
		+ (P^-\zeta_k(n) + q )^Tz_k \\
		&\qquad \qquad \qquad - \frac{1}{2}\zeta_k(n)^TP^-\zeta_k(n) \big),\\ 
		&\hat{h}_{1,k}: -\qhvac + \zeta_k(n)^TP_cz_k-\frac{1}{2}\zeta_k(n)^TP_c\zeta_k(n)=0. 
		\end{aligned}
		\end{align}
		\textbf{Solve for $z^*$:} 
		\begin{align}\label{eq:opt_cvx}
		\begin{aligned}
		z^* = &\arg \min_{z_k|_{k=0}^{\tp-1}} \quad \hat{J}  \\
		\text{ s. t. }  
		& \text{equality constraints}~\eqref{eq:qhvac_hat},~\eqref{eq:opt_ss1}-\eqref{eq:opt_tma} \\
		& \text{inequality constraints}~\eqref{eq:opt_mdot}-\eqref{eq:opt_tzslack}\\
		& k=0,\dots,\tp-1
		\end{aligned}
		\end{align} \\
		\textbf{Update iteration:} Set $n\leftarrow n+1, \; \zeta(n) \leftarrow z^*$.
	}	
	\KwOut{$z^*\leftarrow \zeta(n)$}
\end{algorithm} 

\begin{proposition}~\cite{zeng2020autonomous}\label{prop:descent_cvx_feasible}
	Problem~\eqref{eq:opt_cvx} is feasible and convex, and Algorithm~\ref{algo:convexCtrl} is a descent and convergent algorithm.
\end{proposition}

\begin{remark}
	Proposition~\ref{prop:descent_cvx_feasible} guarantees reliable performance of Algorithm~\ref{algo:convexCtrl}. Since problem~\eqref{eq:opt_cvx} is feasible and convex, if the algorithm converges within the allowable time, it converges to a local minimum of the original non-convex problem. If the algorithm must be stopped before convergence due to inadequate time, the solution obtained has a lower cost than solutions from previous iterates since it is a descent algorithm.
\end{remark}

\subsubsection{Choice of convex approximation method}\label{subsubsec:convexify_choice}
Apart from the convex-concave procedure we used,  there are many approximation methods for non-convex optimization problems that involve bilinearities. The commonly used methods are (i) Branch-and-Bound (BnB)~\cite{mccormick1976computability}  and (ii) Alternate Convex Search (ACS)~\cite{wendell1976minimization}. Next we show that these methods are not applicable to our problem~\eqref{eq:opt_org}, leaving CCP as the only candidate. The following two propositions will be needed for that discussion. 
\begin{proposition}~\cite{zeng2020autonomous}\label{prop:dual_unbounded}
	The dual of problem of~\eqref{eq:opt_org} is unbounded from below.
\end{proposition}
\begin{proposition}\label{prop:soln_boundary}
	Every solution of Problem~\eqref{eq:opt_org} is a boundary solution.
\end{proposition}
The proof of Proposition~\ref{prop:soln_boundary} is provided in the Appendix.

\paragraph{Inapplicability of Branch-and-Bound (BnB)} BnB requires construction of a tight convex under-estimator of the NLP within any given region of the space of the variables~\cite{mccormick1976computability}. The most widely-used under-estimators are Lagrangian relaxation~\cite{d2003relaxations} or convex relaxation. However Proposition~\ref{prop:dual_unbounded} shows the dual of our problem~\eqref{eq:opt_org} is unbounded from below. Therefore Lagrangian relaxation cannot be applied. For convex relaxation, common options are McCormick envelope~\cite{mccormick1976computability} and Reformulation Linearization Technique (RLT)~\cite{sherali1992new}. Both of them reformulate a problem via the addition of certain nonlinear constraints that are generated by using the products of the bounding constraints. However, constructing such products require knowledge of bounds on variables that are involved. In our problem, thermal comfort limits do not have known bounds because of the introduction of slack variables. Hence convex relaxation is also not applicable for our case.

\paragraph{Inapplicability of Alternate Convex Search (ACS)}
ACS~\cite{wendell1976minimization} divides variable set into disjoint blocks and in every step, only the variables of an active block are optimized while those of the other blocks are fixed. Analyses and examples from~\cite{gorski2007biconvex,floudas1990global} show that this method will most likely fail to find a local optimum for problem with boundary solutions (our case). Only initial guesses that belong to a particular set will converge to a local optimum. Because there is no guarantee on convergence to local minima, we do not use ACS. 

\section{(Block I) ID and Prediction}\label{sec:ID+prediction}
\subsection{Identification}
The job of the identification block of Fig.~\ref{fig:MPCflowBrief} is to use data to identify parameters of the zone temperature dynamics model~\eqref{eq:thermal_mdl}, along with the unmeasurable occupant-induced disturbance. We rewrite the model in a different form to describe the identification problem precisely:
\begin{align}\label{eq:thermal_mdl_id}
\begin{split}  
x_{k+1} &= A x_k + B^{id}u^{id}_k + F^{id}\bar{w}_k\\
y_k (=\tz_k )  &= C x_k + D^{id}u^{id}_k + G^{id}\bar{w}_k
\end{split}
\end{align}
Here the state $x_k\in\R^2$, the output $y_k \in \R$, and the matrices $A,C$ are the same as in~\eqref{eq:thermal_mdl}. But while the four inputs in~\eqref{eq:thermal_mdl_id} were divided into controllable and not controllable, here they are divided into measurable and non-measurable. In particular, $u^{id}_k : = [\qhvac,\toa,\solirr]_k \in R^3$ consists of the measurable inputs to the thermal dynamics and the transformed disturbance $\bar{w}_k \in \R$ is the non-measurable input. Other than the regrouping, the two models are identical. Among the three components of $u^{id}_k$,  \qhvac\ is computed from measurements of \mdot\ and \tsa\ using~\eqref{eq:qhvac_def}, and the remaining two inputs can be obtained from a weather station. The output $\tz$ is measured with a sensor.

The system identification algorithm used here is the SPIDR method proposed in our earlier work~\cite{ZengSimultaneousAutomatica:2018}. Fix $i$ as the current time when system identification is to be carried out. Define $\tau_i := \{i-N,i-N+1,\dots, i-1\}$ and $(u^{id},y)_{j}, j \in  \tau_i$ be the measured input-output data for the model~\eqref{eq:thermal_mdl_id} over that time interval. The algorithm SPIDR takes this data and produces an estimate of the model parameters $\mathcal{M}:=(A,B^{id},F^{id},C,D^{id},G^{id})$ and an estimate of the transformed disturbance $\bar{w}_{j}, j \in  \tau_i$. We denote these estimates $\hat{\mathcal{M}}_i$ and $\hat{\bar{w}}_{j}, j \in  \tau_i$ since they depend on $i$. The SPIDR algorithm is executed at time instants $i, i+N_{ad}, i+2N_{ad},\dots$, with $N_{ad}$ large so that enough time has after the previous identification to warrant updating the estimates of the model and disturbance.

The SPIDR algorithm comes with the following guarantees~\cite{ZengSimultaneousAutomatica:2018}:
\begin{enumerate}
	\item The computation involved in obtaining the estimates (model and disturbance signal) is a feasible and convex optimization problem with a strictly convex cost.
	\item The model $\hat{\mathcal{M}}_i$ is BIBO stable and has a positive DC gain from each of the three measurable inputs (outdoor temperature, solar irradiance, and HVAC heat injection) to indoor temperature.
	\item There is exactly one parameter that requires tuning by a human expert. This tuning can be done once (one data set). The two properties mentioned above hold irrespective of the value of this parameter. 
\end{enumerate}

\begin{remark}\label{rem:autonomy}
	The first property ensures that the the system identification algorithm can be executed periodically without any human intervention, i.e., \emph{autonomously}. Autonomy is also helped by the third feature. The second feature helps in two ways. One, it ensures that the model identified is consistent with the physics of HVAC systems. Two, it helps in state estimation. At every decision instant $i$, a Kalman filter is used to estimate the state of the thermal model~\eqref{eq:thermal_mdl}, which is then used as the initial state by the MPC planner : $\hat{x}_0$ in~\eqref{eq:opt_ss1}. The stability guarantee of the model mentioned above ensures that the the Kalman filter is stable~\cite{rhodes1971tutorial}. 
\end{remark}

\subsection{Forecasts of uncontrollable inputs}\label{subsec:forecast-disturbance}
Two types of uncontrollable inputs appear in the thermal model~\eqref{eq:thermal_mdl}, and thus their forecasts over the planning horizon is needed by the MPC planner: weather variables and transformed disturbance $\bar{w}$. These forecasts are obtained as follows.
\begin{enumerate}
	\item Weather variables: Obtain forecast of $[\toa,\solirr]_k^T$ over the next planning horizon from an online weather service.
	\item Transformed disturbance $\bar{w}$: If the prediction horizon does not contain a holiday, assign the disturbance for the same time interval from the previous week estimated by the system identifier, as the forecast. If the prediction horizon contains a holiday, use the disturbance estimate from the same time interval of the previous Saturday as the forecast.
\end{enumerate}

\subsection{Putting them all together}\label{subsec:ctrl_arch}
The components described so far are now combined to form the proposed controller. The architecture is described in Algorithm~\ref{algo:autoMPC}. Recall that Figure~\ref{fig:MPCflowBrief} shows the complete closed loop system. 
\begin{algorithm}[h] \label{algo:autoMPC}
	\LinesNotNumbered
	\caption{Proposed MPC architecture}
	\SetAlgoLined
	\KwIn{planning horizon $\tp\in\Z^+$, 
		control horizon $\tc\in\Z^+$,  
		and model updating interval $N_{ad}\in\Z^+$. } 	
	\textbf{Setup:} $\Set_c\eqdef\{\tc,2\tc,\dots\}$, $\Set_{ad}\eqdef\{N_{ad},2N_{ad},\dots\}$.\\
	\For{$i=1,2,\dots$}{
		\If{$i\in\Set_{ad}$}{
			\textbf{Measure:} Input $u^{id}$ and output $y$ of the model~\eqref{eq:thermal_mdl_id}, over the time interval $[i-N_{ad}:i-1]$.\\
			\textbf{System ID:} Estimate model $\hat{\mathcal{M}}_i$ and disturbance $\hat{\bar{w}}[i-N_{ad}:i-1]$ using the SPDIR algorithm from~\cite{ZengSimultaneousAutomatica:2018}.
		}
		
		\textbf{Estimate state:} Estimate current state $\hat{x}[i]$ of thermal model~\eqref{eq:thermal_mdl} using a Kalman filter.\\
		\textbf{Predict disturbance:} As described in Section~\ref{subsec:forecast-disturbance}.\\
		\textbf{Optimize:} Compute control decisions $\mdot[i:i+\tp-1]$ and $\tsa[i:i+\tp-1]$ using Algorithm~\ref{algo:convexCtrl}. \\
		\textbf{Implement:} Apply $\mdot[i:i+\tc-1]$ and $\tsa[i:i+\tc-1]$ to the plant.\\
	}
\end{algorithm}





\subsection{Baseline controller for comparison}\label{subsec:baseline}
The baseline controller is chosen to be the single-maximum controller which is widely used in practice~\cite{ASHRAE_handbook_fund:09}. The single-maximum controller operates the HVAC system in three modes depending on where the zone temperature $\tz$ is compared to the deadband $[\tzmin, \tzmax]$. When $\tz$ exceeds the upper bound $\tzmax$, reheat is turned off and the supply airflow rate $\mdot$ is increased with the help of a PI controller. When the zone temperature is below the lower bound $\tzmin$, the airflow rate $\mdot$ is kept at the minimum allowed value but the supply air temperature is increased with the help of a PI controller. When the zone temperature is in the deadband $[\tzmin, \tzmax]$, the supply air temperature is kept at $\tca$ and the flow rate are both kept at the minimum allowed value.

\section{Simulation setup}\label{sec:setup}
To assess performance of the proposed control system, we perform closed loop simulations for nearly a year-long period with a realistic time varying plant. Simulations with the baseline controller are also performed on the same plant for comparison. The plant model on which the controller acts is calibrated to mimic a large auditorium in a building in the University of Florida campus (Pugh Hall). The auditorium in Pugh Hall is served by an air handling unit, and it has the same HVAC system configuration as shown in Figure~\ref{fig:HVAC}.

\subsection{Plant description}\label{sebsec:plant}
The plant is a time varying non-linear ordinary differential equation, and with a large internal heat load \dist\ (kW). 
\begin{equation*}
\resizebox{1\hsize}{!}{$
	\begin{aligned}
	& C_z(t)\dot{T}^z(t) = \frac{\tw(t)-\tz(t)}{R_z(t)} +\qhvac(t)  + \Ae(t)\solirr(t)+\dist(t) \\ 
	& C_w(t)\dot{T}^w(t) = \frac{\toa(t)-\tw(t)}{R_w(t)}+\frac{\tz(t)-\tw(t)}{R_z(t)}
	\end{aligned}      
	$}
\end{equation*}
where \tw\ (\degree C) is the wall temperature, $C_z(t)$, $C_w(t)$, $R_z(t)$, $R_w(t)$ are the time-varying thermal capacitances and resistances of the zone and wall, respectively, and \Ae(t)\ is the effective area of the building for incident solar radiation. One can view this model as a  time-varying version of the commonly used RC-network models of building thermal dynamics.

\begin{figure}[htb]
	\centering
	\includegraphics[width=1\columnwidth]{\figPATH/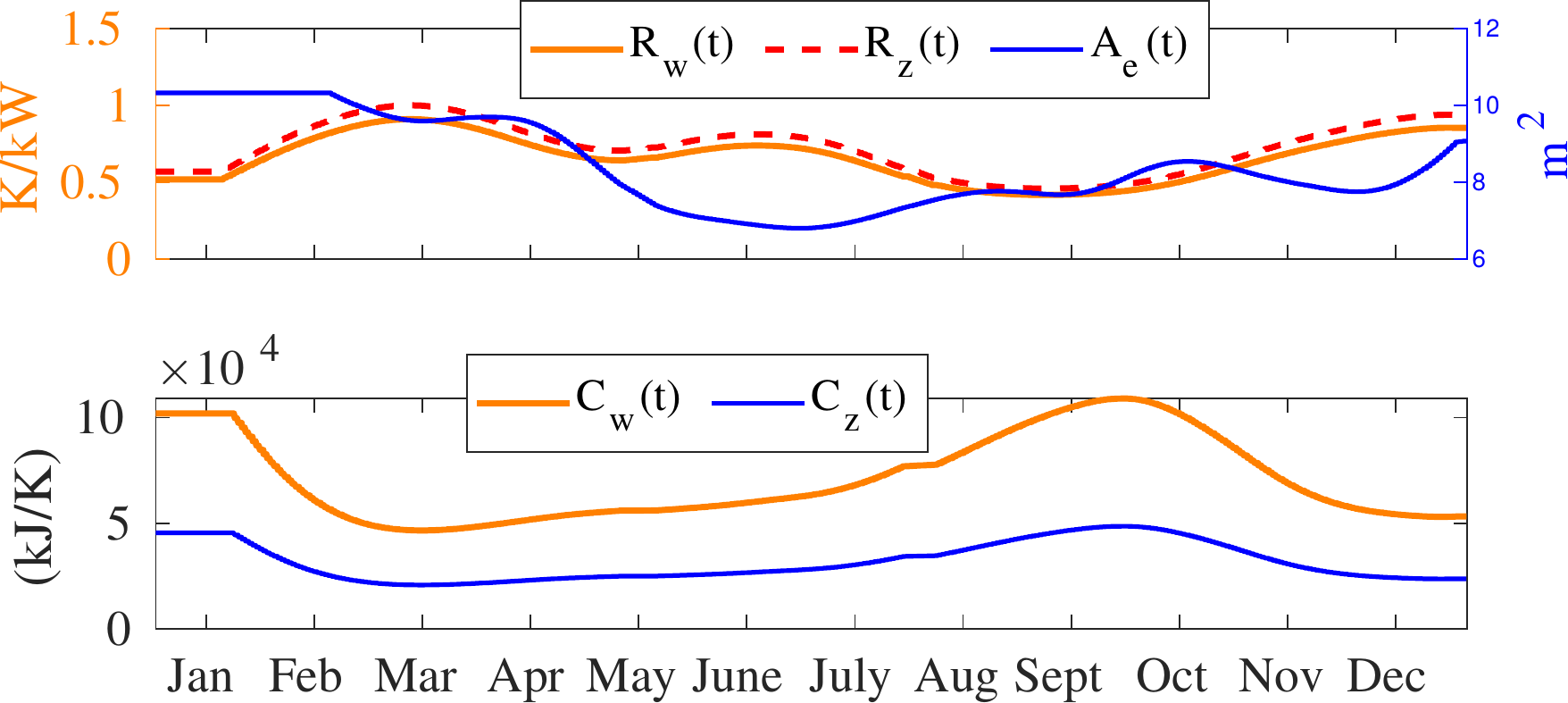}
	\caption{Time-varying parameters of the plant. }
	\label{fig:para_plant}
\end{figure}

The time-varying plant parameters are shown in Fig.~\ref{fig:para_plant}, which are chosen as follows. The average values of the time-varying parameters are chosen to be the same as the values given in~\cite{coffman2018simultaneous}~[Table~2], which contains the plant parameters estimated using data from an auditorium in Pugh Hall located in the University of Florida. 

\subsection{Closed loop parameters}\label{subsec:sim_para}
The planning horizon for MPC is 1 day and the control horizon is $15$ minutes, with a sampling time $\Delta t =5$ minutes, so $\tp=288$ and $\tc=3$. The total time span for MPC is 50 weeks. The number of decision variables for problems~\eqref{eq:opt_org} and~\eqref{eq:opt_cvx} is $2592$($=9\tp$).

Thermal comfort and flow rate constraints depend on whether the building is in occupied or unoccupied mode~\cite{ashrae55:2017}. The maximum occupancy for Pugh Hall auditorium is 229 persons, and its occupied mode (occ) is scheduled from 6:30 AM to 10:30 PM while the remaining time is deemed unoccupied (unocc). We used these parameters for the simulation. The thermal comfort bounds are $[21.9,23.6]\degree$C for occupied mode, and $[21.1,24.4]\degree$C for unoccupied mode. The minimum allowed value for the supply airflow rate $\mdot^{min}$ is computed based on the ventilation requirements specified in ASHRAE 62.1~\cite{ASHRAE62-2016}. More specifically, $\mdot^{min,unocc}$ is computed assuming 0 occupancy for the unoccupied period with 31\% occupancy for the occupied period. Note for the baseline controller, $\mdot^{min,occ}$ is kept as high as $1.90$ kg/s, otherwise the baseline controller fails to maintain the zone temperature comfort satisfactorily. The remaining parameters are listed in Table~\ref{tab:para_constraint}.
\begin{table*}[htb]
	\centering
	\caption{Parameters for baseline and MPC controllers.}
	\label{tab:para_constraint}
	\begin{threeparttable}
		\begin{tabular}{l l l l | l l l | l l l}  \hline
			& \multicolumn{1}{l}{Unoccupied} & \multicolumn{1}{l}{Occupied}  &  & $\tsamin$ & $12.8$ & \degree C  & $\tca$  & $12.8$ & \degree C  \\ 
			$\tzmin$  & $21.1$ & $21.9$ & \degree C & $\tsamax$ & $37.8$ & \degree C & $\cop$     & $3.5$  & N/A     \\ 
			$\tzmax$  & $24.4$ & $23.6$ & \degree C & $\tsarate$ & $0.56$ & \degree C/min & $\alpha$   & $0.3$  & N/A  \\ 		
			$\mdot^{min}$& $0.95$ & $1.47$, $1.90$ \tnote{*} & kg/s & $\mdot^{max}$& $4.72$  & kg/s  & $a_f$ & $417.5$& W/(kg/s)$^2$   \\ 
			& &  &      &   $\mdot^{rate}$& $0.2$  & kg/s/min  &  & &  \\ 	\hline
		\end{tabular}
		\begin{tablenotes}
			\item[*] $\mdot^{min,occ}=1.47$ is used for the MPC controllers, and $\mdot^{min,occ}=1.90$ is used for the baseline controller.
		\end{tablenotes}	
	\end{threeparttable}	
\end{table*} 

\begin{figure}[htb]
	\centering
	\includegraphics[width=1\columnwidth]{\figPATH/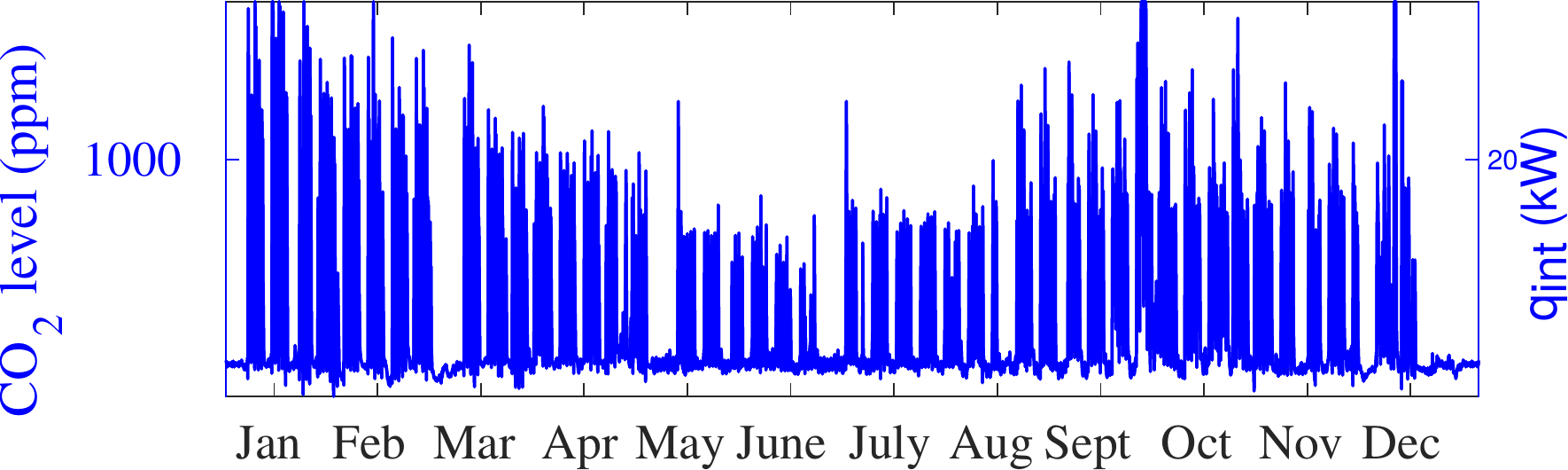}
	\caption{(Left) \cotwo\ level measurements from the auditorium in Pugh Hall during the year of 2013. (Right) The internal heat load \dist.}
	\label{fig:dist4ctrl}
\end{figure}

The uncontrollable input signals are chosen as follows: solar irradiance data \solirr\ is taken from NSRDB: \url{https://nsrdb.nrel.gov/}, and ambient temperature \toa\ is taken from \url{weatherunderground.com}, both for Gainesville, FL from the year of 2013. The internal heat load (\dist) is chosen by scaling \cotwo\ data collected from the auditorium in Pugh Hall during the same year, which is shown in Figure~\ref{fig:dist4ctrl}. The high resolution and long term data collection was made possible by using a custom made data logger~\cite{middelkoop2020an}. The rationale is that occupancy is correlated to the \cotwo\ level. Note that the heat load is by design a large, time-varying, and aperiodic signal. 


All numerical results presented in this work are obtained through \MATLAB. Specifically, the plant is simulated in \simulink. The system identification problem from~\cite{ZengSimultaneousAutomatica:2018} for estimating model and disturbance is solved using \cvx~\cite{cvx} package. For control computation, the nominal non-convex problem~\eqref{eq:opt_org} is solved using \ipout~\cite{wachter2006implementation} package, and the proposed convex problem~\eqref{eq:opt_cvx} is solved using \cvx~\cite{cvx} package. We used a desktop computer with a $3.60$GHz $\times$ $8$ CPU and $16$ GB RAM,  running Linux, for the closed loop simulations.



\section{Simulation results}\label{sec:results}
A total of five distinct controllers are tested through simulations on the same plant:
\begin{itemize}
	\item \textbf{Baseline:} the single-max controller described in Section~\ref{subsec:baseline}.
	\item \textbf{Proposed (Adapt-CVX):} the proposed controller (Algorithm~\ref{algo:autoMPC}), with both model update and convex planner for control computation.
	\item \textbf{NAdapt-CVX:} the proposed controller (Algorithm~\ref{algo:autoMPC}), \emph{but without updating the dynamic model and the disturbance estimates}. 
	\item \textbf{Adapt-NCVX:} the proposed controller (Algorithm~\ref{algo:autoMPC}), \emph{but using the non-convex problem~\eqref{eq:opt_org} instead of the convex problem from Algorithm~\ref{algo:convexCtrl} to compute commands.} 
	\item \textbf{NAdapt-NCVX:} MPC with the nominal non-convex problem~\eqref{eq:opt_org} for computing control commands, and without updating the dynamic model and the disturbance estimates. Note this the MPC architecture generally described in the literature. 
\end{itemize}
In all the controllers that uses a non-convex optimization, if the NLP solver is unable to converge before the control update interval is over, decisions computed by the baseline controller are sent to the actuators. 

\subsection{Comparison with the baseline controller}\label{subsec:cmp_baseline}
\begin{figure*}[htb]
	\centering
	\includegraphics[width=1\linewidth]{\figPATH/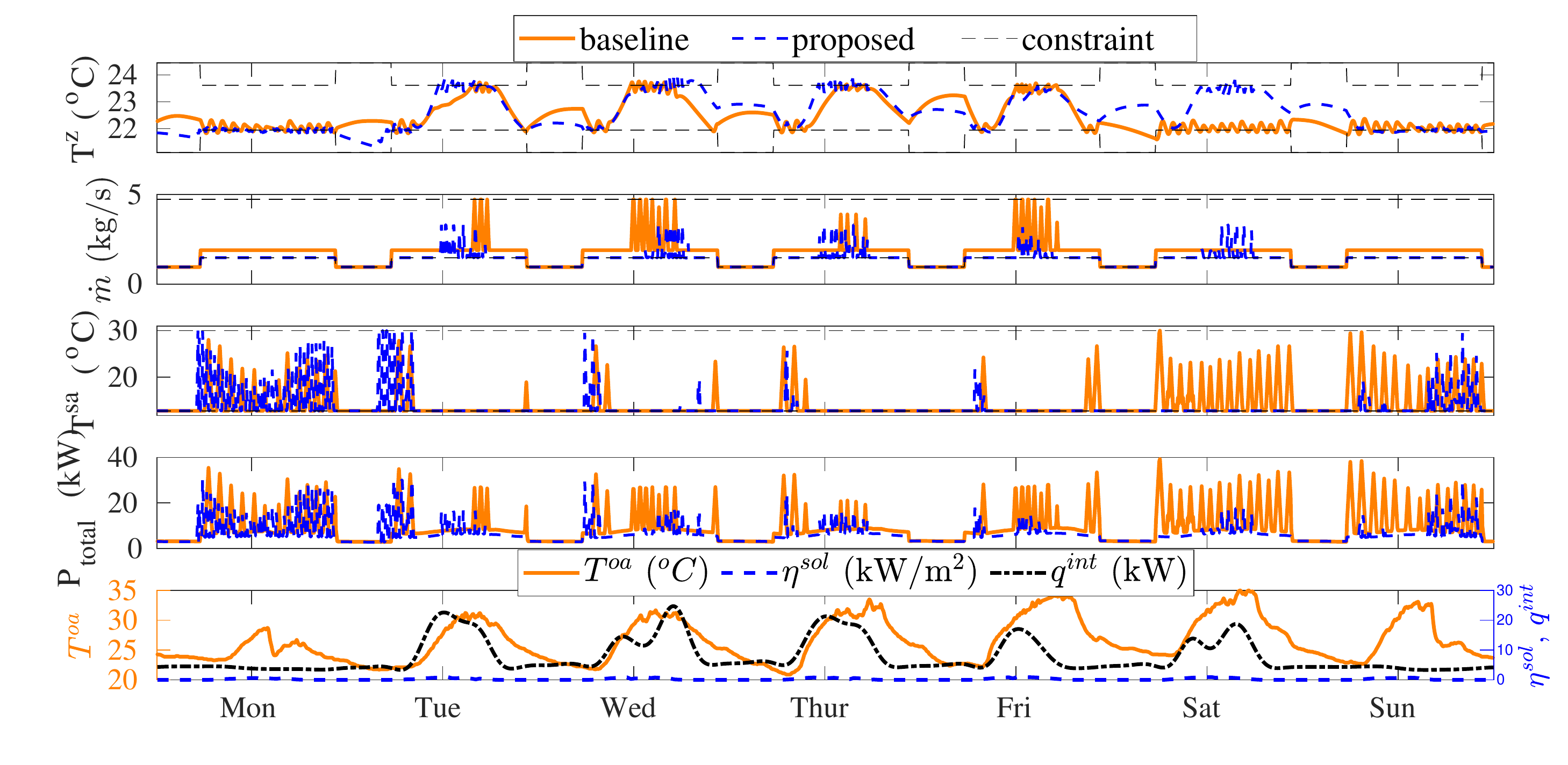}
	\caption{Comparison of the simulation results for the proposed MPC scheme vs baseline controller, during Aug/26/2013 - Sept/1/2013. }
	\label{fig:decisionCMP_MPC_base_zoom}
\end{figure*}

\begin{figure*}[htb]
	\centering
	\includegraphics[width=1\linewidth]{\figPATH/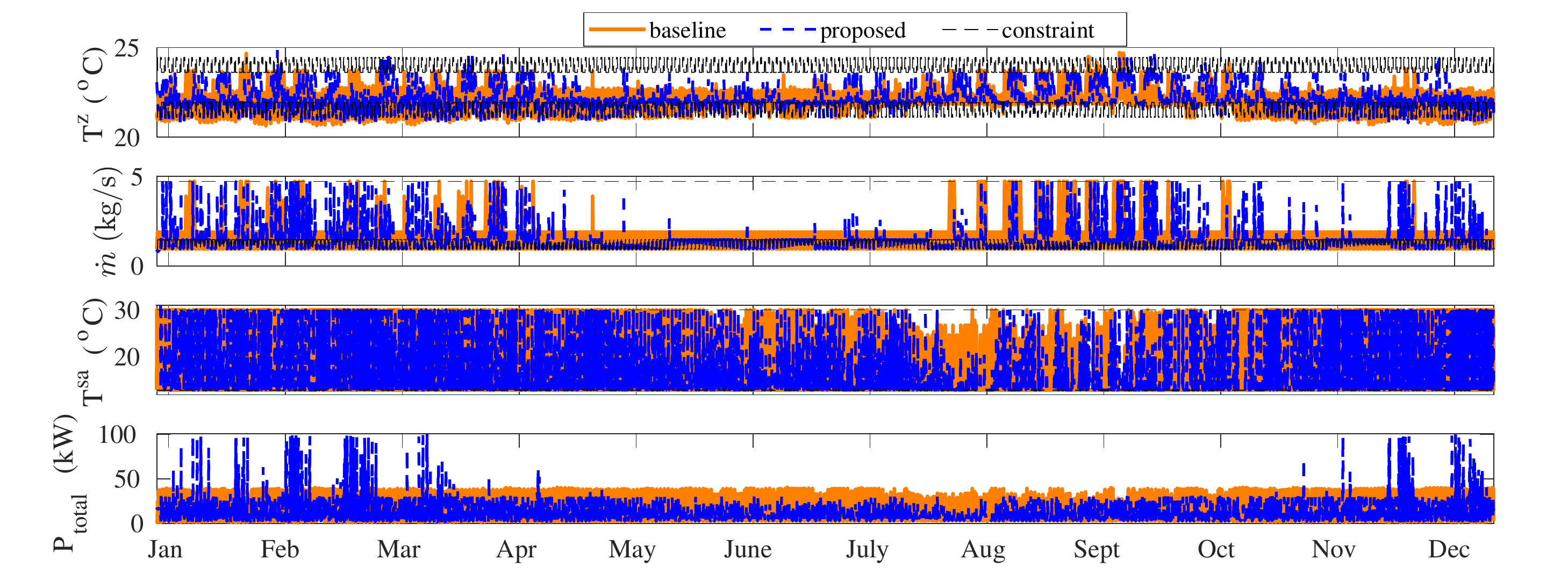}
	\caption{Comparison of the simulation results for the proposed MPC scheme vs baseline controller. }
	\label{fig:decisionCMP_MPC_base}
\end{figure*}

The proposed MPC scheme outperforms the baseline controller in both maintaining zone temperature and reducing energy use; see Table~\ref{tab:ctrl_cmp}. Data on uncontrollable inputs, control command, and the output (zone temperature) are shown in Figure~\ref{fig:decisionCMP_MPC_base_zoom} for the full 50 weeks.  Figure~\ref{fig:decisionCMP_MPC_base} zooms into one week: Aug/26/2013 - Sept/1/2013.

In particular, the proposed controller (Algorithm~\ref{algo:autoMPC}) reduces energy use by $26.8\%$ over the baseline controller, to EUI = $53.5$ kBtu/(ft$^2\cdot$year); see Table~\ref{tab:ctrl_cmp}. The baseline controller is already more efficient than the average controller in the field: its site EUI for the tested period is $72.9$ kBtu/(ft$^2\cdot$year), which is lower than the median site EUI = $84.3$ kBtu/(ft$^2\cdot$year) for college buildings in the United States~\cite{star2018portfolio}. 

The improvement in performance over the baseline controller is consistent with results in the literature that  have compared MPC with baseline controllers.  MPC's ability to use disturbance forecasts and prediction from the model allows it to make better decisions than a purely output feedback controller.

\subsection{Benefit and necessity of the design features}
\begin{table*}[htb]
	\centering
	\caption{Performance comparison among various controllers.}
	\label{tab:ctrl_cmp}
	\resizebox{1\hsize}{!}{
	\begin{tabular}{l | l  l  l  l}\hline
		controller  & \multicolumn{1}{l}{site EUI (kBtu/(ft$^2\cdot$yr))} & \multicolumn{1}{l}{planner failure (\%)}  & \multicolumn{1}{l}{RMSE of \tz\ violation (\degree C)}  & \multicolumn{1}{l}{ Max \tz\ violation (\degree C)}\\ \hline
		Baseline     & 72.9 &  N/A  & 0.45  & 2.3\\ 
		NAdapt-NCVX  & 63.4 &  0.4  & 0.46  & 4.0  \\ 		
		NAdapt-CVX   & 63.9 &  0    & 0.41  & 1.7 \\ 
		Adapt-NCVX   & 53.7 &  0.1  & 0.22  & 3.2 \\ 
		Adapt-CVX (Proposed)    & 53.5 &  0    & 0.23  & 1.1\\
		\hline
	\end{tabular}
	}
\end{table*}

\paragraph{Need for model and disturbance update:}\label{subsec:cmp_nonadaptMPC}
We tested the role and/or value of adaptation by turning off the adaptation block. A model and disturbance (for a week) are estimated from data from the first week of 2013. They are used by the controller for every week of the year. The resulting MPC controller is referred to by the ``NAdapt-'' prefix, e.g., in Table~\ref{tab:ctrl_cmp}. We see from the table that adaptation reduces energy use by about 16\% and reduces zone temperature violation over the non-adaptation case.

Thus, adaptation - periodically updating models and disturbances from data - is both necessary (for indoor comfort) and beneficial (improves energy use) for an MPC-based controller for HVAC systems.

\begin{figure}[htb]
	\centering
 	\includegraphics[width=1.05\linewidth]{\figPATH/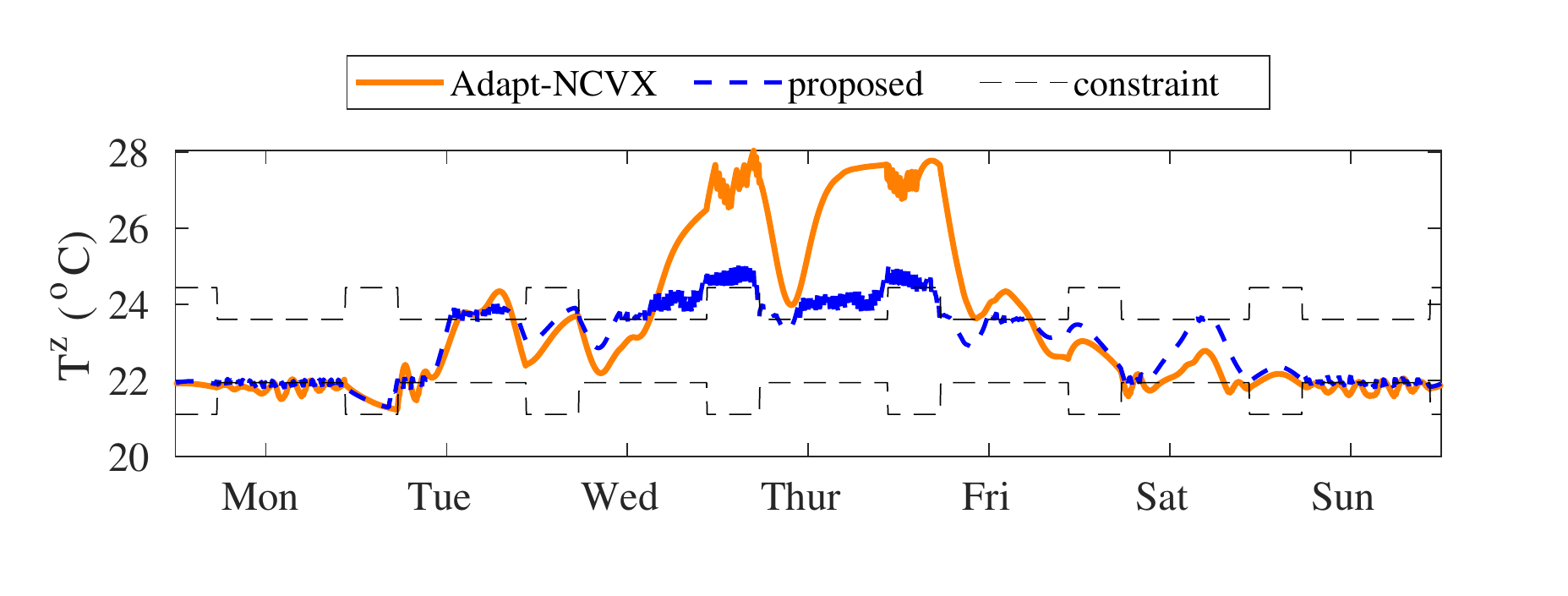}
	\caption{Comparison of the simulation results of the zone temperature for the proposed MPC scheme vs Adapt-NCVX controller, during Sept/30/2013 - Oct/6/2013. }
	\label{fig:decisionCMP_MPC_NCVX}
\end{figure}

\paragraph{Need for convexification of the MPC planner:}\label{subsec:cmp_noncvxMPC} 
NLP solvers such as Ipout~\cite{wachter2006implementation} are quite powerful. Thus, solving the non-convex MPC planning problem~\eqref{eq:opt_org} is \emph{usually} not an issue. On average it takes about $2.7$ seconds for Ipopt to find a local minimum of the non-convex problem, failing to do so with the available 15 minutes \emph{only $0.1\%$} of the time. When this happens, decision from the baseline controller is used as control commands. The resulting switching control action can lead to large violation in the indoor temperature. See Fig.~\ref{fig:decisionCMP_MPC_NCVX} for an example of this phenomenon. The zone temperature exceeds the upper bound by 3.2 \degree C for an extended period of time. Thus, though a non-convex planner rarely fails, when it does it leads a catastrophic loss of performance that will render the control system unacceptable to the user.

In contrast to MPC with a non-convex planner, the proposed MPC scheme with a convexified planner \emph{always} finds a minimum within the available 15 minutes, taking $1.7$ seconds on average to compute the control decisions. Partially as a result of that, \emph{it is able to provide the best performance in maintaining zone temperature among all five controllers tested.}

Therefore, even though solving the nominal non-convex problem is rarely an issue, in those rare occasions the controller can cause serious disruption to occupant's thermal comfort.  It is unlikely such a control system will be acceptable to building owners and occupants. In short, the convex approximation of the MPC planner is necessary. 

It should be noted that the NAdapt-NCVX controller is the MPC scheme generally used in the literature, e.g.~\cite{afram2014theory,Foucquier2013state,Oldewurtel2012use,RamanMPC_AE:2020}. Without the benefits from both of the designed features, this controller has a maximum zone temperature violation of $4.0\degree$ C (even though it occurs rarely) and does not perform as well as the proposed controller in terms of energy use.

\begin{remark}
	We remark here the performance delivered by the proposed MPC scheme is obtained under strong plant-model mismatch in the following aspects: (i) The plant is time-varying and nonlinear, while the MPC planner uses a linear model. (ii) The proposed MPC scheme assumes both the plant and the disturbance are the same as that from the previous week, but the plant and the disturbance do not satisfy those properties.
\end{remark}

\section{Conclusion}\label{sec:conclusion}
This paper takes a first stab at designing an MPC-based control system for HVAC systems that can operate autonomously for long periods, without requiring intervention of human experts. Autonomy is made possible by two features: (i) automated periodic update of thermal model and internal disturbance signals, and (ii) a convex approximation of the MPC planner's optimization problem. The year long simulations shows that both of the features are essential to get the performance improvement over the simple baseline controller over a long period. The need for periodic re-learning the model and disturbance is easy to see in the context of buildings. The need for convexity in the planning problem is less obvious at the design stage, but was discovered from the simulation results. Even though the nominal non-convex planning problem can be used effectively nearly 100\% of times, the rare instances it fails to converge causes dramatic fluctuations in the indoor temperature rendering the control system an unlikely contender for real-life application. Without these features, though MPC can outperform the baseline controller in certain scenarios, the benefits may not be substantial enough to defray the additional cost of implementing MPC. 

At the current stage the proposed MPC architecture uses arguably one of the simplest schemes for forecasting of the internal disturbance. It is envisioned that a more accurate prediction scheme, possibly with the aid of technologies such as occupancy recognition or \cotwo\ level sensing, should further improve performance of the MPC controller.

Many extensions of this work are possible. The most immediate next step is extending the proposed control scheme to include humidity dynamics and ventilation requirements, which will require including as part of the control commands the outdoor airflow and conditioned air temperature (downstream of the cooling/dehumidification coil; see Fig.~\ref{fig:HVAC}). These two have been assumed fixed in this paper but in fact can be commanded through the building automation system. It should be possible to reduce energy use even more and provide better thermal comfort by including outdoor airflow and conditioned air temperature into the list of control commands. The challenge is to incorporate the nonlinear humidity dynamics in zone thermal models and the nonlinear process models of the cooling/dehumidification coil~\cite{RamanMPC_AE:2020}. The autonomy achieved by the control system proposed  here is due to the use of linear dynamic models. 
Other useful directions include extension to multi-zone buildings, improvements in the forecasting methodology for the internal disturbance, etc. 


%
% ...

\bibliographystyle{asmems4}

\bibliography{\TZbibPATH/Tingting,\DiCEbibPATH/Barooah,\DiCEbibPATH/optimization,\DiCEbibPATH/systemid,\DiCEbibPATH/grid,\DiCEbibPATH/building-vx,\DiCEbibPATH/sensnet_bib_dbase}

\appendix       
\section*{Appendix}\label{sec:appendix}
\begin{proof}[Proof of Proposition~\ref{prop:opt_org_feasible}]
	It suffices to find one feasible solution to Problem~\eqref{eq:opt_org}. Let ${\mdot_k}^*=\mdot^{min}$ and ${\tsa_k}^*=\tsamin$, $\forall k=0,\dots,\tp-1$, which satisfy the actuator constraints~\eqref{eq:opt_mdot}-\eqref{eq:tsa_rate2}. Values of ${\tma_k}^*,{\tz_k}^*,{\qhvac_k}^*,{x^T_{k+1}}^*$ are dependent on $({\mdot_k}^*,{\tsa_k}^*)$, and are solved from the set of linear independent equations~\eqref{eq:opt_qhvac}-\eqref{eq:opt_tma}, $\forall k=0,\dots,\tp-1$. Based on the resulting ${\tz_k}^*$, where $k=0,\dots,\tp-1$, it is straightforward to show that ${\epsilon^{min}_k}^*=max\{0,\tzmin-{\tz_k}^*\}$ and ${\epsilon^{max}_k}^*=max\{0,{\tz_k}^*-\tzmax\}$ are the corresponding minimizers to Problem~\eqref{eq:opt_org}, which also satisfy constraints~\eqref{eq:opt_tz1}-\eqref{eq:opt_tzslack}. Therefore, we found $z^*=[{z^*_0}^T,\dots,{z^*_k}^T,\dots,{z^*_{\tp-1}}^T]^T$, where $z_k^*= [{\mdot_k}^*,{\tsa_k}^*,{\tma_k}^*,{\tz_k}^*,{\qhvac_k}^*,{x^T_{k+1}}^*,{\epsilon^{min}_k}^*,{\epsilon^{max}_k}^*]^T \in \R^{9}$, as one feasible solution to Problem~\eqref{eq:opt_org}.
\end{proof}

\begin{proof}[Proof of Proposition~\ref{prop:soln_boundary}]
	We show this by contradiction. 
	
	Assume $z^*=[z^*_0,\dots,z^*_k,\dots,z^*_{\tp-1}]^T$ is an interior optimal solution to Problem~\eqref{eq:opt_org}, we show $z^*$ does not satisfy the KKT conditions. 
	
	The Lagrangian of~\eqref{eq:opt_org} is:
	\begin{align}\label{eq:opt_org_lagrang}
	\begin{aligned}
	\Lagr(z,\lambda,\upsilon) &= \sum_{k=0}^{\tp-1}\big( \frac{1}{2}z_k^TPz_k+q^Tz_k+\\
	&\qquad \sum_{p=1}^{5}\lambda_{p,k}h_{p,k}+\sum_{q=1}^{12}\upsilon_{q,k}f_{q,k}\big),
	\end{aligned}
	\end{align}
	where $h_{1,k} - h_{5,k}$ denotes the equality constraints~\eqref{eq:opt_qhvac}-\eqref{eq:opt_tma}, and $f_{1,k} - f_{12,k}$ denotes the inequality constraints~\eqref{eq:opt_mdot}-\eqref{eq:opt_tzslack}, $\forall k=0,\dots,\tp-1$, respectively.	
	
	For an interior point $z^*$, inequality constraints are inactive at $z^*$, which implies $\upsilon_{q,k} =0$, $\forall q=1,\dots,12,k=0,\dots,\tp-1$. Since $z^*$ is optimal, we have for Lagrangian~\eqref{eq:opt_org_lagrang},
	\begin{align}
	\begin{aligned}\label{KKT_cond}
	0& = \frac{\partial \Lagr}{\partial z_k}|_{z_k=z^*_k}, \forall k=0,\dots,\tp-1\\ 
	\implies 0& = Pz^*_k +q+\sum_{p}\lambda_{p,k}\frac{\partial h_{p,k}}{\partial z_k}|_{z_k=z^*_k}\\
	& = (P+\lambda_{1,k}P_c)z^*_k+q+
	\begin{bmatrix}
	0 \\ 0 \\ -\lambda_{5,k}\\ \vdots
	\end{bmatrix}.
	\end{aligned}
	\end{align}
	It suffices to find one contradiction that the set of equations~\eqref{KKT_cond} is not possible. Expand the second entry of Equation~\eqref{KKT_cond}, one writes
	\begin{align}
	\cpa(1+\lambda_{1,k})\mdot^*_k = 0 \implies \lambda_{1,k}=-1,
	\end{align}
	because $\mdot\geq\mdot^{min}>0$. Substitute $\lambda_{1,k}=-1$ into the first entry from~\eqref{KKT_cond} we have
	\begin{align*}
	2\alpha_f\mdot_k^*+\frac{\cpa}{\cop}{\tma_k}^*+\cpa{\tz_k}^* = -\cpa\tca\frac{1+\cop}{\cop},
	\end{align*}
	which cannot hold since LHS$>$0 whereas RHS$<$0. Therefore we show any interior point does not satisfy the KKT condition, meaning Problem~\eqref{eq:opt_org} only has boundary solutions.      
\end{proof}

%

\end{document}